\documentstyle[12pt]{ioplppt}
\input axodraw1.sty

\def\be{\begin{equation}}
\def\ee{\end{equation}}
\def\beq{\begin{eqnarray}}
\def\eeq{\end{eqnarray}}
\def\s{\sigma} 
\def\G{\Gamma} 
\def\F{_2F_1}
\def\an{analytic}
\def\ac{\an{} continuation}

\def\ndim{NDIM}
\def\half{\frac{1}{2}}
\def\threehalf{\frac{3}{2}}

\begin{document}
\jl{1}
\title{Two-loop self-energy diagrams worked out with NDIM}  
\author{Alfredo T. Suzuki\footnote{E-mail: suzuki@axp.ift.unesp.br} and Alexandre G. M.
Schmidt\footnote {E-mail: schmidt@power.ift.unesp.br}}
\address{Instituto de F\'isica Te\'orica, Universidade Estadual
Paulista, R.Pamplona 145, S\~ao Paulo  SP, CEP 01405-900, Brazil. }

\begin{abstract}
In this work we calculate two two-loop massless Feynman integrals pertaining to
self-energy diagrams using NDIM (Negative Dimensional Integration Method). We
show that the answer we get is 36-fold degenerate. We then consider special
cases of exponents for propagators and the outcoming results compared with
known ones obtained via traditional methods.
\end{abstract}
\maketitle

\section{Introduction.}

The dimensionality of space-time plays a key role in all branches of Physics.
The quantities we calculate, as theoreticians, depend very much on the 
number of dimensions we are considering. Theories in higher and lower
dimensions than four have been put forth by many researchers and plentiful of
good insights have been gained through this exercise. Zooming in the arena of
quantum field theory, we discover that the dimensionality of space-time gained
a more sophisticated status, being promoted from a mere integer number to that
of a complex variable, with the advent and development of dimensional
regularization by 't Hooft {\it et al} and several other pioneers in the
field \cite{thooft}.

In other words, we could say that quantum field theory (QFT), besides other
great ideas it inspired, physical and mathematical alike, did reveal this
amazing possibility: the \ac{} of the space-time dimension $D$.

The union between the theory of analytic functions and QFT is very profitable.
Dimensional regularization (DREG), the technique that bears the concept of
analytically continued $D$, is one of its profits. As a step further in this
direction Halliday {\it et al} \cite {halliday, halliday2} developed the idea
of analytically continued $D$ to {\it negative values}. Of course, the seminal
idea of negative values for $D$ is already contained in the work of 't Hooft
and others. But what is novel in Halliday's insight is the amazing possibility
of letting field propagators be raised to positive powers, so that the integrand
becomes polynomial. The thrust behind the idea is that solving a polynomial
integral should be --- in principle at least --- easier to perform than
rational ones elicited in the usual Feynman integrals. This very simple
argument, which we call negative dimensional integration method (\ndim{}), can
simplify the calculation of Feynman integrals in an astounding way
\cite{suzuki1,suzuki2,suzuki3,box,lab,boxnew,2loops}.

In the usual DREG \cite{collins,nash,wilson} the only quantities that preserve
their meaning are the Green's functions \cite{faddeev}. We will not try to
discuss whether they still have (or have not) their meaning preserved within
the context of \ndim{} nor speculate what are the features, if any, of this
``new world'' of negative dimensions \cite{lab}. What we do is simply to allow
for it just for calculational purposes. The reader must have this important
point in mind.

In our previous works \cite{box,lab} we calculated massive one-loop four point
functions (former reference) and a massless two-loop three-point vertex (latter
reference) with the \ndim{} approach. In the first, \ndim{} provided not only
the well-known hypergeometric functions but six other new results in a very
straightforward manner \cite{boxnew}; while for the two-loop vertex, we
considered the particular case where two of its external momenta were set
on-shell, and \ndim{} responded with as many as twelve times --- surprisingly
enough, all of them yielding the same correct result --- that is, a twelvefold
degeneracy. This led us to conjecture that when the power series had unit
argument and they were all summable, then the result would be degenerate. That
is, if this conjecture is correct, we need only to carry out one sum --- the
most convenient one, of course. The conjecture remains to be proven or
disproved.

Here we put our \ndim{} to another ``lab-test'' \cite{lab} by considering two
two-loop self-energy diagrams which we call by the funny name ``flying saucer''
diagrams --- side view (Fig. 1) and front view (Fig. 2), just to make it easier
for us to refer to them. The outline for this article is as follows: In Section
2 we solve the two-loop Feynman integral relative to these two graphs, i.e.,
the space-time dimension and the exponents of the pertinent propagators are
left arbitrary. Then, in Section 3, we particularize to suit either the
``flying saucer, side view'' or the ``flying saucer, front view'' diagram
cases. And finally, Section 4 is devoted to our concluding remarks.

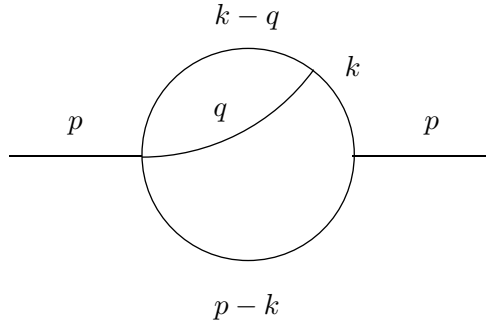
\begin{figure}
\begin{center}
\vspace{15mm}
\begin{picture}(180,110)(0,80)
\thinlines
\BCirc(110,140){40}
\CArc(70,219)(80,-90,-36)
\put(150,140){\line(1,0){50}}
\put(70,140){\line(-1,0){50}}
\small{
\put(110,80){\makebox(0,0)[b]{$p-k$}}
\put(110,190){\makebox(0,0)[b]{$k-q$}}
\put(180,150){\makebox(0,0)[b]{$p$}}
\put(45,150){\makebox(0,0)[b]{$p$}} 
\put(100,155){\makebox(0,0)[b]{$q$}} 
\put(150,170){\makebox(0,0)[b]{$k$}} }
\end{picture}
\end{center}
\caption{Two-loop massless Feynman diagram: the ``flying saucer'', side view.} 
\end{figure}

\section{Feynman Graphs with Four Massless Propagators.}
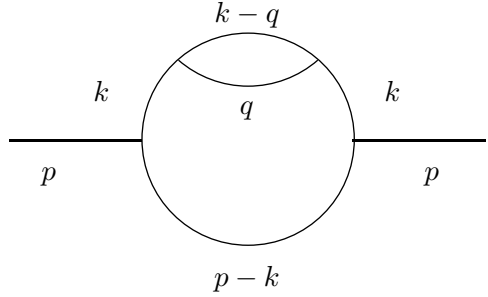
\begin{figure}
\begin{center}
\vspace{20mm}
\begin{picture}(180,110)(0,80)
\thinlines
\BCirc(110,140){40}
\CArc(110,200)(40,-131,-49) 
\put(150,140){\line(1,0){50}}
\put(70,140){\line(-1,0){50}}
\small{
\put(110,85){\makebox(0,0)[b]{$p-k$}}
\put(110,185){\makebox(0,0)[b]{$k-q$}}
\put(180,125){\makebox(0,0)[b]{$p$}}
\put(35,125){\makebox(0,0)[b]{$p$}} 
\put(110,150){\makebox(0,0)[b]{$q$}} 
\put(165,155){\makebox(0,0)[b]{$k$}}
\put(55,155){\makebox(0,0)[b]{$k$}} }
\end{picture}
\end{center}
\caption{Two-loop massless Feynman diagram: the ``flying saucer'', front view.} 
\end{figure}

The \ndim{} approach to solve Feynman integrals is beautiful in its simplicity:
First, we take the propagators of the integral we want to work out, multiply
each one of them by a specific parameter and then solve the $D$-dimensional
gaussian integral whose argument is that very expression. Let us see how it
works in practice. Consider the gaussian integral,
 
\be \label{gauss} I(p^2;D) = \!\!\int\! \d^D\!k\;\d^D\!q \exp{\left[
-\alpha q^2\!-\beta k^2-\! \gamma(p-k)^2-\!\omega(k-q)^2 \right]}, \ee   
which clearly is relevant to the diagrams we want to work out. It is not
difficult to integrate it; the result is,
\be \label{taylor} I(p^2;D) = \left(\frac{\pi^2}{\lambda}
\right)^{D/2} \exp{\left[-\frac{\gamma p^2}{\lambda} 
(\beta\omega+\alpha\beta+\alpha\omega)\right]},\ee 
where
$\lambda=\alpha\beta+\beta\omega+\alpha\gamma+\gamma\omega+\alpha\omega$.
Expanding (\ref{taylor}) in Taylor series and also expanding the multinomial
expression in $\lambda$, we get an eightfold summation,
\beq \label{serie} I(p^2;D) &=& \sum_{n_i=0}^\infty
\frac{(-p^2)^{n_1+n_2+n_3}
(-n_1-n_2-n_3-\half
D)!}{n_1!n_2!n_3!n_4!n_5!n_6!n_7!n_8!}\nonumber\\ 
&& \times\alpha^{n_1+n_2+n_4+n_6+n_8}\beta^{n_1+n_3+n_4+n_5}
\gamma^{n_1+n_2+n_3+n_6+n_7} \nonumber\\
&&\times\omega^{n_2+n_3+n_5+n_7+n_8},\eeq 
with the constraint $-n_1-n_2-n_3-\half D=n_4+n_5+n_6+n_7+n_8$ coming from the
multinomial expansion.

The second step is simpler and faster:  expand the exponential
(\ref{gauss}) in Taylor series first to get,
\be \label{ndim} I(p^2;D) = \sum_{i,j,l,m=0}^\infty (-1)^{i+j+l+m}
\frac{\alpha^i\beta^j\gamma^l\omega^m}{i!j!l!m!}{\cal J}(i,j,l,m;D),\ee 
where we define,
\be {\cal J}(i,j,l,m;D) = \int \d^D\!k\;\d^D\!q\;
(q^2)^i(k^2)^j\left[(p-k)^2\right]^l
\left[(k-q)^2\right]^m ,\ee 
which is our negative dimensional integral.

Comparing (\ref{serie}) and (\ref{ndim}) we get an expression for the
negative-$D$ integral,

\be \label{geral} {\cal J}(i,j,l,m;D) = (-\pi)^D(p^2)^\s\;G
\sum_{n_i=0}^\infty \frac{1}{n_1!n_2!n_3!n_4!n_5!n_6!n_7!n_8!} ,\ee 
where we define the product of gamma functions,
\[ G = \G(1+i)\G(1+j)\G(1+l)\G(1+m)\G(1-\s-\half D) ,\]  
and since the two expressions must equal, sum indices in the former
and exponents of propagators in the latter, must satisfy the system,
\be
\left\{\begin{array}{rcr}
n_1+n_2+n_4+n_6+n_8&=&i\\
    n_1+n_3+n_4+n_5&=&j\\
n_1+n_2+n_3+n_6+n_7&=&l\\
n_2+n_3+n_5+n_7+n_8&=&m\\
        n_1+n_2+n_3&=&\s
        \end{array} \right.    
\ee
where $\s=i+j+l+m+D$ and the last equation comes from the multinomial
expansion. Observe that the equations above are linear, but because we have
eight unknowns and only five equations, in order to solve this system we must
choose three of the unknowns and solve it in terms of them. There are many ways
in which this choice can be done; in fact, there are $C^8_3=8!/(5!3!)=56$
possibilities altogether. However, $20$ out of the $56$ lead us to trivial
solutions, which present no interest at all. The remaining $36$ give us the
results for the Feynman integral when we plug their solutions in equation
(\ref{geral}). 

We will solve the non-trivial systems and write down the general results, but
before doing that, let us see what we can do to lessen our task. Looking at the
Feynman diagram we can spot symmetry properties that help us in this. Thus,
we expect the outcoming result to be symmetric under the exchange $i
\leftrightarrow m$, which in turn will reduce by half the number of distinct
systems that we need to deal with, since the symmetry will account for the
remaining half.

Let us then first consider the solution that leaves $n_5,\;n_6,\;n_8$ as free
indices in the summation; let us call it ${\cal J}_a$. It yields, 
\beq {\cal J}_a &=& (-\pi)^D(p^2)^\s P_1\sum_{n_5,n_6,n_8=0}^\infty
\frac{(-1)^{n_6}
(-i-m-\half D|n_8)(-l+\s|n_6)}{n_5!n_6!n_8!}
\nonumber\\
&&\times \frac{(-j-m-\half D|n_5-n_6)(\half D+l|n_5+n_8)}
{(1+l-m|n_5-n_6+n_8)}, \eeq 
where 
\beq P_1 &=& \frac{\G(1+i)\G(1+j)\G(1+l)\G(1+m)}
{\G(1+l-m)\G(1+i+m+\half D)\G(1+j+m+\half D)\G(1+l-\s)}\nonumber\\
&&\times \frac{\G(1-\s-\half D)}{\G(1-l-\half D)} \nonumber,
\eeq  
with the Pochhammer symbol \cite{lebedev} denoted by, 
\[ (a|m)\equiv (a)_m = \frac{\G(a+m)}{\G(a)} .\] 

Using one of the properties of the Pochhammer symbol \cite{lebedev}, i.e.,
\be \label{AC}(a|-m) = \frac{(-1)^m}{(1-a|m)} ,\ee 
one can identify these series as hypergeometric \cite{bailey}; in
fact, we can rewrite them in a convenient manner using another property, 
\be \label{agrupar}(a|b+c) = (a|b)(a+b|c) ,\ee 
and sum, for example, the $n_8$ series using the well-known
formula \cite{lebedev},  
\be \label{f21} \F(a,b;c|1) = \frac{\G(c)\G(c-a-b)}{\G(c-a)\G(c-b)},\ee
yielding,
\beq {\cal J}_a &=& (-\pi)^D(p^2)^\s P_1P_2\sum_{n_5,n_6=0}^\infty
\frac{(1+i|-n_6)(-j-m-\half D|n_5-n_6)}{n_5!n_6!(1-m-\half D|-n_6)}
\nonumber\\
&&\times\frac{(\half D+l|n_5)(-l+\s|n_6)}{(1+i+l+\half D|n_5-n_6)},\eeq 
where 
\be P_2 = \frac{\G(1+i)\G(1+l-m)}{\G(1-m-\half D)\G(1+i+l+\half
D)}.\ee 

In a similar manner we can sum the two remaining series, getting as a result,
\beq \label{final} {\cal J}_a &=& (-\pi)^D(p^2)^\s\frac{\G(1+i)\G(1+l)
\G(1+m) \G(1-\s-\half D)}{\G(1+\s)\G(1-i-\half D)\G(1-m-\half D)
\G(1-l-\half D)} \nonumber\\
&& \times \frac{\G(1+i+j+m+\half D)\G(1-i-m-D)}{\G(1+i+m+\half
D)\G(1+l-\s)}.\eeq 

The last and final step that need to be taken is to bring this result back to
our real physical world with positive $D$. Grouping the gamma functions in the
numerator with the ones in the denominator in convenient Pochhammer symbols and
using (\ref{AC}), we arrive at
\beq \label{correto} &{\cal J}&_a^{AC} = \pi^D (p^2)^\s (-i|i+m+\half
D)(-m|i+m+\half D)(\s+\half D|-2\s-\half D)\nonumber\\
&&\times (-l|\s)(-i-j-m-\half D|j)(i+m+D|-i+l-m-\half D) .\eeq 

This very simple operation allows us to analytically continue the result back
into our real physical world, $D>0$. Equation (\ref{correto}) is the general
result, and we note that it is symmetric in $i \leftrightarrow m$ as it should
be, and it is correct \cite{narison}. The reader will ask {\em immediately}:
What is(are) the result(s) that the other solution(s) provide? \ndim{} answers
in as brief and surprising a manner as it could be: {\em the same}. Indeed
just to make sure we went through all of them, and verified that it is possible
to sum all the emerging series and they provide the same result, namely,
equation (\ref{final}) which leads to the correct expression, equation
(\ref{correto}), i.e., we have a thirty-six-fold degeneracy!

A word of caution here: Not all the sums can be so easily dealt with. Yet, just
to convince the reader that it is possible to sum them all, we shall carry out
one more summation, the hardest one. The degeneracy above mentioned can be
classified into two sets: $32$ solutions are like ${\cal J}_a$ with relatively
easy sums to carry out, while $4$ of them are like the following one which we
call ${\cal J}_b$. Consider then the solution with indices $n_1,\;n_4,\;n_5$,
 
\beq {\cal J}_b &=& (-\pi)^D (p^2)^\s P_3\sum_{n_1,n_4,n_5=0}^\infty 
\frac{(-1)^{n_4}(\half D+l|n_4+n_5)(\half D+m|n_1+n_4)}
{n_1!n_4!n_5!(1-j+\s|n_4+n_5)} 
\nonumber\\
&&\times \frac{(-j|n_1+n_4+n_5)}{(1-i-j-\half D|n_1+n_4)} ,\eeq
where
\be P_3 = \frac{\G(1+i)\G(1+l)\G(1+m)\G(1-\s-\half D)}
{\G(1-j+\s)\G(1-m-\half D)\G(1-i-j-\half D)\G(1-l-\half D)}.\ee

Using the same procedure we used for ${\cal J}_a$, we can carry out the $n_5$
summation, yielding,
\beq {\cal J}_b &=& (-\pi)^D (p^2)^\s P_3 P_4\sum_{n_1,n_4=0}^\infty 
\frac{(\half D+l|n_4)(\half D+m|n_1+n_4)(-\s|n_1)}
{n_1!n_4!(\half D+l-\s|n_1+n_4)} 
\nonumber\\
&&\times \frac{(-j|n_1+n_4)}{(1-i-j-\half D|n_1+n_4)} ,\eeq 
where
\[ P_4 = \frac{\G(1-j+\s)\G(1-l-\half D+\s)}{\G(1+\s)\G(1-j-l-\half
D+\s)}.\]

However, this time, neither of the remaining sums (in $n_1$ or $n_4$) can be
written in terms of $_2F_1$, but rather in terms of $_3F_2$ which are not
summable for arbitrary values of its parameters. Here comes the trick that will
do the required job: Put the $n_4$ series in terms of a $_3F_2$ function and
use the property \cite{bailey}
\be _3F_2( a,b,c;e,f|1) = Q\;_3F_2(e-a,\;f-a,\;s;\;\;s+b,\;s+c|1) ,\ee
where $s=e+f-a-b-c$ and
\[Q=\frac{\G(e)\G(f)\G(s)}{\G(a)\G(s+b)\G(s+c)}.\]

A good choice is to take
\be
\begin{array}{lcl}
a&=&-j+n_1 \\
b&=&\half D+l \\
c&=&\half D+m+n_1\\
e&=&\half D+l-\s+n_1\\
f&=&1-i-j-\half D+n_1
\end{array}
\ee
so that the gamma functions in $Q$ simplify several factors in the series and
some of them can be grouped by the property (\ref{agrupar}) giving,
\beq {\cal J}_b &=& (-\pi)^D (p^2)^\s P_3 P_4 P_5 \nonumber\\
&&\times \sum_{n_1,n_4=0}^\infty \frac{(\half D+j+l-\s|n_4)(\half D+m|n_1)(-\s|n_1)}
{n_1!n_4!(1-i-\s-\half D|n_1+n_4)} 
\nonumber\\
&&\times \frac{(1-i-\half D|n_4)(1-i-m-\s-D|n_4)}
{(1-i+l-m-\s-\half D|n_4)} ,\eeq 
where
\[ P_5 = \frac{\G(\half D+l-\s)\G(1-i-j-\half D)\G(1-i-m-\s-D)}
{\G(-j)\G(1-i-\s-\half D)\G(1-i+l-m-\s-\half D)} .\]

Now the sum in $n_1$ is a $\F$ function and we can sum it using the Gauss
summation formula (\ref{f21}). The gamma functions that arise from this summation simplify
the remaining series in $n_4$ from a $_3F_2$ into a $\F$ function that can be
summed again using the usual Gauss summation formula. The end result is,
\be {\cal J}_b = (-\pi)^D (p^2)^\s P_3 P_4 P_5 P_6 P_7,\ee
where $P_6$ came from the $n_1$ sum,
\[ P_6 = \frac{\G(1-i-\s-\half D)\G(1-i-m-D)}{\G(1-i-\half
D)\G(1-i-m-\s-D)} ,\] 
and $P_7$ came from the last one,
\[ P_7 = \frac{\G(-j)\G(1-i+l-m-\s-\half D)}{\G(\half
D+l-\s)\G(1-\s+l)}.\]  

Multiplying all the gamma factors ($P_3,...,P_7$) we get, exactly, the
expression (\ref{final}), that leads to the correct result
(\ref{correto}). 

\section{Special Cases.}

The scalar integral we calculated in the previous section has particular cases
of interest, namely, the ``flying saucer'' diagrams, side view (Fig.1) and
front view (Fig.2). For the side view diagram the exponents of the propagators
are all equal to minus one, while for the front view diagram the exponents
are minus one except for $j=-2$.

Let us denote by ${\cal J}^{AC}$ the general result for the ``{\em
flying saucer}'' diagram; when we take $i=j=l=m=-1$ we have the result for
the {\em side view} diagram,
\be {\cal J}_{SV}^{AC} = \pi^D(p^2)^{D-4} \frac{\G^3(\half
D-1)\G(D-3)\G(2-\half D)\G(4-D)}{\G(3-\half D)\G(D-2)\G(\threehalf
D-4)} ,\ee
while when we take $i=-1,\;j=-2,\;l=m=-1$ we have the result for the front
view diagram,
\be {\cal J}_{FV}^{AC} = \pi^D(p^2)^{D-5} \frac{\G^3(\half D-1)\G(5-D)
\G(D-4)\G(2-\half D)}{\G(4-\half D)\G(D-2)\G(\threehalf D-5)} ,\ee 
which reproduce well-known results \cite{narison,muta}.

\section{Conclusion.}

The \ac{} of the space-time dimension $D$ into {\em negative values} has shown
us advantages never dreamed of before: we interpret the \ac{} like in usual
DREG but solve the Feynman integrals in a much simpler way because the
integrands we have to deal with are polynomial. The way back road, via another
\ac{} is straightforward and the whole procedure is quite simple and elegant.
There are no cumbersome parametric integrals to solve; on the contrary, the
only things one needs to know are how to solve gaussian integrals, and systems
of linear algebraic equations! Furthermore, we have
surprising manifold degenerate solutions for a single integral. Our previous
conjecture on this topic seems to hold, though further research is needed to
prove it.

\begin{ack}
AGMS wishes to thank CNPq (Conselho Nacional de Desenvolvimento
Cient\'ifico e Tecnol\'ogico) for financial support.
\end{ack}

\end{document}